\documentstyle[prd,aps,eqsecnum,preprint,tighten,floats]{revtex}
\begin{document}
\draft

\preprint{\vbox{\hfill OHSTPY--HEP--T--96--anp97.11.20}
          \vbox{\vskip0.5in}
          }

\title{Matrix Theories from Reduced $SU(N)$\\
 Yang-Mills with Adjoint Fermions }

\author{Francesco Antonuccio\footnote{address starting December 1996 
Max Planck Institute, Heidelberg, Germany} and Stephen S. Pinsky}
\address{Department of Physics, The Ohio State University, Columbus,
OH 43210}

\date{\today}

\maketitle

\begin{abstract} We consider a dimensional reduction
of 3+1 dimensional $SU(N)$ Yang-Mills theory coupled to adjoint
fermions to obtain a class of $1+1$ dimensional matrix field
theories. We derive the quantized
light-cone Hamiltonian in the light-cone gauge $A_- = 0$ 
and large-$N$ limit, 
and then solve for
the masses, wavefunctions and structure functions of the color singlet
``meson-like'' and ``baryon-like'' boundstates. Among the states 
we study are many massless string-like states that can be solved for 
exactly.
\end{abstract}
\pacs{ }

\section{Introduction}
Recently it has been conjectured that a non-perturbative 
formulation of M theory
may be realized as a large-$N$  supersymmetric matrix model
\cite{baf96}  formulated in the light-cone Hamiltonian approach. 
In this paper we discuss and solve a class of $1+1$ dimensional
matrix field theories 
that are obtained from a similar dimensional reduction
procedure applied to $3+1$ dimensional Yang-Mills theory. 
While our motivation stems from QCD, the non-perturbative 
techniques and results (both analytical and numerical) of this
work are expected
to be of interest in the context of these new developments.

A key strategy in formulating such model field theories
 is to retain as many of the essential degrees of freedom
of higher dimensional QCD while still being able to extract 
complete non-perturbative solutions. In this present work,
we start by considering
${\mbox{QCD}}_{3+1}$  coupled to Dirac adjoint fermions.
Here, the virtual creation of fermion-antifermion
pairs is not suppressed in the large-$N$ limit -- in contrast 
to the case for fermions in the fundamental representation 
\cite{and96b} -- and so  
one may study the structure of boundstates beyond the 
valence quark (or quenched) approximation. 
We also anticipate that the techniques employed here 
will have special interest in the context
of solving supersymmetric matrix theories.

The ${\mbox{QCD}}_{3+1}$  theory coupled to adjoint fermions may be
reduced to a $1+1$ dimensional field theory by stipulating that all
fields are independent of the transverse coordinates $x^{\perp}=(x^1,x^2)$.
The resulting theory is 
${\mbox{QCD}}_{1+1}$  coupled to two $1+1$ dimensional complex adjoint
spinor fields, and two
real adjoint scalars. One also finds Yukawa interactions between
the scalars and fermion fields. 
While this approach is not equivalent to solving the full 
$3+1$ theory
and then going to the regime where $k_\perp$ is relatively small, it may share
many qualitative features of the higher dimensional theory,
since the longitudinal dynamics is treated exactly. Studies
of this type for pure glue and with fundamental 
quarks have yielded
a number of interesting results \cite{and96a,and96b,buv95}.

The unique features of light-front quantization
\cite{dir49} make it a potentially powerful tool for the nonperturbative
study of
quantum field theories.  The main advantage of this approach is the apparent
simplicity of the vacuum state.  Indeed, naive kinematic arguments
suggest that
the physical vacuum is trivial on the light front. Since in this
case all fields
transform in the adjoint representation of $SU(N)$, 
the gauge group of the theory
is actually $SU(N)/Z_N$, which has 
nontrivial topology and vacuum structure. For the particular gauge group
 $SU(2)$ this has been discussed elsewhere \cite{pir96}.
While this vacuum structure may in fact be relevant for a discussion on
condensates, for the purposes of this calculation they will be ignored.

In the first section we formulate the $3+1$ dimensional $SU(N)$ Yang-Mills
theory and then perform dimensional reduction to obtain
a $1+1$ dimensional matrix field theory.
The light-cone Hamiltonian is then derived for the
light-cone gauge $A_- = 0$ following a discussion
of the physical degrees of freedom of the theory. 
Singularities from Coulomb interactions are regularized in a
natural way, and we outline how particular ``ladder-relations''
take care of potentially troubling singularities for
vanishing longitudinal momenta $k^+ = 0$.
A short section on exact massless solutions 
of the boundstate integral equations 
is given. In the final section we discuss
and present the numerical results of our work, including 
mass spectra, evidence for phase transitions
at certain critical couplings, and the behavior of
polarized and unpolarized structure functions for 
the lightest meson-like and
baryon-like boundstates.

\section{SU(N) Yang-Mills Coupled to Adjoint Fermions: Definitions}

We first consider $3+1$ dimensional $SU(N)$ Yang-Mills coupled to
a Dirac spinor field whose components transform in the adjoint 
representation of $SU(N)$:
\begin{equation}
{\cal L} = \mbox{Tr} \left[ -\frac{1}{4} F_{\mu \nu} F^{\mu \nu} 
 + \frac{{\rm i}}{2} 
(\bar\Psi\gamma ^{\mu} \buildrel \leftrightarrow \over
D_{\mu}\Psi) - m\bar\Psi \Psi \right] \; ,
\label{3+1theory}
\end{equation}
where $D_{\mu} = \partial_{\mu} + {\rm i}g [ A_\mu,\ \ ] $ and $F_{\mu \nu}
= \partial_{\mu}A_{\nu} - \partial_{\nu} A_{\mu} + {\rm i}g [A_{\mu},
A_{\nu} ]$. We also write $A_{\mu} = A_{\mu}^a \tau ^a$ where $\tau^a$ is
normalized such
that $\mbox{Tr} (\tau^a \tau^b ) = \delta_{ab}$.
The projection operators\footnote{
We use the conventions $\gamma^{\pm} = (\gamma^0 \pm \gamma^3)/
\sqrt{2}$, and $x^{\pm}=(x^0 \pm x^3)/\sqrt{2}$.} 
$\Lambda_L,\Lambda_R$ permit a decomposition
of the spinor field $\Psi = \Psi_L +\Psi_R$, where
\begin{equation}
\Lambda_L = {1 \over 2} \gamma^+ \gamma^- ,\quad \Lambda_R = {1 \over 2}
\gamma^- \gamma^+ \quad \mbox{and} \quad
\Psi_L = \Lambda_L \Psi, \quad \Psi_R = \Lambda_R \Psi .
\end{equation}
Inverting the equation of motion for $\Psi_L$, we find
\begin{equation}
\Psi_L ={1 \over 2{\rm i}D_-} \left[ {\rm i} 
\gamma^i D_i +m \right ] \gamma^+ \Psi_R
\label{eqnmotion}
\end{equation}
where $i=1,2$ runs over transverse space. Therefore $\Psi_L $ is not an
independent degree of freedom. 

Dimensional reduction of the $3+1$ dimensional Lagrangian 
(\ref{3+1theory}) is performed by assuming (at the classical level)
that all fields are independent of the transverse coordinates
$x^{\perp}=(x^1,x^2)$: $\partial_{\perp} A_{\mu} =0$ and 
$\partial_{\perp} \Psi = 0$.
In the resulting $1+1$ dimensional field theory, the transverse components
$A_{\perp} = (A_1,A_2)$ of the gluon field will be represented
by the $N \times N$ complex matrix fields $\phi_{\pm}$:
\begin{equation}
     \phi_{\pm} = \frac{A_1 \mp {\rm i} A_2}{\sqrt{2}}.
\end{equation}   
Here,  $\phi_-$ is just the Hermitian conjugate of 
$\phi_+$. When the theory is quantized, $\phi_{\pm}$ will correspond to
$\pm 1$ helicity bosons (respectively). 

The components of the Dirac spinor $\Psi$ are the $N \times N$
{\em complex} matrices $u_{\pm}$ and $v_{\pm}$, which 
are related to the left and right-moving spinor fields according
to 
\begin{equation}
\Psi_R ={1 \over 2^{{1 \over 4}}}
\left ( \begin{array}{c} u_+\\ 0 \\ 0 \\u_- \end{array}
\right )
\quad
\Psi_L ={1 \over 2^{{1 \over 4}}}
\left ( \begin{array}{c} 0 \\ v_+ \\ v_- \\0 \end{array}
\right )
\end{equation}
Adopting the light-cone gauge $A_- = 0$ allows one to explicitly
rewrite the left-moving fermion fields $v_{\pm}$ in terms 
of the right-moving fields $u_{\pm}$ and boson fields $\phi_{\pm}$,
by virtue of equation (\ref{eqnmotion}). We may therefore eliminate 
$v_{\pm}$ dependence from the field theory. Moreover, 
Gauss' Law  
\begin{equation}
\partial_-^2 A_+ = g \left( {\rm i}[\phi_+, \partial_- \phi_-] +
                    {\rm i}[\phi_-, \partial_- \phi_+]
 + \{ u_+,u_+^{\dagger} \} + \{ u_-,u_-^{\dagger} \} \right)
\end{equation}
permits one to remove any explicit dependence on $A_+$,
and so the remaining {\em physical}
degrees of freedom of the field theory are represented by
the helicity $\pm \frac{1}{2}$ fermions $u_{\pm}$, and the
helicity $\pm 1$ bosons $\phi_{\pm}$. There are no ghosts
in the quantization scheme adopted here.
In the light-cone frame  
the Poincar\'e generators $P^-$ and $P^+$ for the 
reduced $1+1$ dimensional field theory are given by
\begin{equation}
P^+=\int ^\infty _{-\infty} dx^- \mbox{Tr}  \biggl[
2 \partial_- \phi_- \cdot \partial_- \phi_+
+ {{\rm i} \over 2} \sum_h \left ( u^\dagger_h \cdot 
\partial_- u_h -\partial_-
u^\dagger_h \cdot u_h \right )
\biggr ]
\end{equation}
\begin{equation}
P^- = \int ^\infty _{-\infty} dx^- \mbox{Tr} \biggl[
 m_b^2 \phi_+ \phi_- 
-{ g^2 \over 2}{J}^+ {1\over \partial_-^2 }{J}^+ + 
{ t g^2 \over 2} \left [\phi_+ , \phi_- \right ]^2
+\sum_h{F}^\dagger_h {1 \over {\rm i}\partial_-} {F}^\dagger_h
\biggr ]  
\label{hamiltonian}
\end{equation}
where the sum $\sum_h$ is over $h=\pm$ helicity labels,
and 
\begin{eqnarray}
J^+ & = & {\rm i}[\phi_+, \partial_- \phi_-] +
                    {\rm i}[\phi_-, \partial_- \phi_+]
 + \{ u_+,u_+^{\dagger} \} + \{ u_-,u_-^{\dagger} \} \\
F_{\pm} & = & \mp s g\left[ \phi_{\pm} \; , u_{\mp} \right ] + 
{ m \over \sqrt{2}} u_{\pm} \label{Fterm}
\end{eqnarray}
We have generalized 
the couplings  
by introducing the variables $t$ and $s$, which do not 
spoil the $1+1$ dimensional gauge invariance of the reduced theory;
the variable $t$ will determine the strength of the quartic-like interactions,
and the variable $s$ will determine the 
strength of the Yukawa interactions between the fermion and boson fields,
and appears explicitly in equation (\ref{Fterm}). The dimensional
reduction of the original $3+1$ dimensional theory yields the canonical
values $s=t=1$.

Renormalizability of the reduced theory also requires 
the addition of a 
bare coupling $m_b$, which leaves 
the $1+1$ dimensional gauge invariance intact. 
In all calculations, the renormalized
boson mass ${\tilde m}_b$ will be set to zero.    

Canonical quantization of the field theory
is performed by decomposing the boson
and fermion fields into Fourier expansions
at fixed light-cone time $x^+ = 0$:
\begin{equation}
u_\pm  = {1 \over  \sqrt {2\pi}} \int_{-\infty}^\infty dk
\hspace{1mm} b_\pm(k)  e^{-{\rm i}k x^-} \quad \mbox{and} \quad
\phi_{\pm} = {1 \over  \sqrt {2\pi}} \int_{-\infty}^\infty { dk \over
\sqrt{2|k|}}\hspace{1mm}a_{\pm}(k) e^{-{\rm i}k x^-}
\end{equation}
where $b_{\pm} = b_{\pm}^a\tau^a$ etc.
We also define
\begin{equation}
\quad b_{\pm}(-k) =d_{\mp}^{\dagger}(k),
\quad a_\pm(-k) = a_{\mp}^\dagger (k),
 \label{note}
\end{equation}
where $d_{\pm}$
correspond to antifermions. 
Note that  
$(b^{\dagger}_{\pm})_{ij}$
should be distinguished
from $b_{\pm ij}^{\dagger}$, since in the former the quantum conjugate 
operator $\dagger$ acts on (color) indices, while
it does not in the latter. The latter formalism
is sometimes customary in the study of matrix models. 
The precise connection between the usual gauge theory and matrix theory
formalism may be stated as follows:
\[
 b_{\pm ji}^{\dagger} =
b_{\pm}^{a\dagger}\tau^{a*}_{ji}=b_{\pm}^{a\dagger}\tau^a_{ij} = 
(b_{\pm}^{\dagger})_{ij}
\]
The commutation and anti-commutation relations (in matrix formalism)
for the boson and fermion fields take the following
form in the large-$N$ limit ($k,{\tilde k} >0$; $h,h'={\pm}$): 
\begin{equation}
\left [ a_{h ij}(k),a_{h' kl}^{\dagger}({\tilde k}) \right] =
\{ b_{h ij}(k), b_{h' kl}^{\dagger}({\tilde k}) \}
= \{ d_{h ij}(k), d_{h' kl}^{\dagger}({\tilde k}) \}
= \delta _{h h'} \delta_{jl}\delta_{ik} \delta(k-{\tilde k}),
\label{rhccrs}
\end{equation}
where have used the relation
$\tau^a_{ij} \tau^a_{kl} =
\delta_{il} \delta_{jk} -{1 \over N} \delta_{ij} \delta_{kl}$. 
All other (anti)commutators vanish.

The Fock space of physical states 
is generated by the color singlet states, which have a natural
`closed-string' interpretation. They are formed by  
a color trace  of the fermion, antifermion and boson
operators 
acting on the vacuum state $|0\rangle$.
Multiple string states couple to the theory with
strength $1/N$,
and so may be ignored.

\section{The Light-Cone Hamiltonian}
For the special case ${\tilde m}_b=m=t=s=0$, the light-cone Hamiltonian
is simply given by the current-current term $J^+ \frac{1}{\partial_-^2} J^+$
in equation (\ref{hamiltonian}). In momentum space,
this Hamiltonian takes the form
\begin{eqnarray}
P^-_{J^+ \cdot J^+} &=& {g^2 \over 2 \pi} 
\int^\infty_{-\infty} dk_1 dk_2 dk_3 dk_4 {\delta(k_1+k_2-k_3-k_4)
\over (k_3-k_1)^2}
{\mbox{Tr} \over 2} \biggl[\nonumber \\
 & & \sum_{h,h'}
:\{b^\dagger_h(k_1),b_h(k_3)\}:
:\{b^\dagger_{h'}(k_2),b_{h'}(k_4)\}: \nonumber \\
&+ &{(k_1+k_3)(k_2+k_4) \over 4 \sqrt{|k_1||k_2||k_3||k_4|}} 
:[a_+^\dagger(k_1),a_+(k_3)]:
:[a_{+}^\dagger(k_2),a_{+}(k_4)]: \nonumber \\
&+ &{(k_2+k_4) \over 2 \sqrt{|k_2||k_4|}} \sum_{h}
:\{ b^\dagger_h(k_1),b_h(k_3)\}: 
:[a_{+}^\dagger(k_2),a_{+}(k_4)]: \nonumber \\
&+ &{(k_3+k_1) \over 2\sqrt{|k_1||k_3|}} \sum_{h'}
:[ a_+^\dagger(k_1),a_+(k_3)]::\{b^\dagger_{h'}(k_2),b_{h'}(k_4)\}:\biggr]
\label{jj}
\end{eqnarray}
The explicit form of the Hamiltonian 
(\ref{jj}) in terms of the operators $b_{\pm}$,
$d_{\pm}$ and $a_{\pm}$ is straightforward to calculate, but too long
to be written down here. It should be stressed, however,
 that several $2 \rightarrow 2$ parton
processes are suppressed by a factor $1/N$, and so are ignored in 
the large-$N$ limit. 
No terms involving $1 \leftrightarrow 3$ parton interactions are 
suppressed in this limit, however.

 One 
can show that this Hamiltonian 
conserves total helicity $h$, which is an additive quantum number.
Moreover, the number of fermions 
{\em minus}
the number of antifermions
is also conserved in each interaction, 
and so we have an additional quantum number ${\cal N}$. 
States with ${\cal N} = 0$ 
will be referred to as {\em meson-like} states, while 
the quantum number  ${\cal N} = 3$ will define {\em baryon-like} states. 

The instantaneous Coulomb interactions involving
$2 \rightarrow 2$ parton interactions behave singularly when
there is a zero exchange of momentum between identical `in'
and `out' states. The same type of
Coulomb singularity
involving $2 \rightarrow 2$ boson-boson interactions
appeared in a much simpler model \cite{dek93}, and can be shown to cancel
a `self-induced' mass term (or self-energy) obtained from normal ordering
the  
Hamiltonian. The same prescription works in the model
studied here.
There are also finite residual terms left over after this cancellation
is explicitly performed for the boson-boson and boson-fermion
interactions, and they cannot be absorbed by a redefinition
of existing coupling constants.   
These residual terms behave as momentum-dependent mass terms,
and in some sense represent the flux-tube energy
between adjacent partons in a color singlet state. 
For the boson-boson and boson-fermion interactions they are 
respectively 
\begin{eqnarray}
\frac{g^2 N}{2 \pi} \cdot {\pi \over 4\sqrt{k_b k_{b'}}} \quad
\mbox{and} \quad
\frac{g^2 N}{2\pi}{1 \over k_f} \left ( \sqrt{1 +{k_f \over k_b}}-1\right)
\end{eqnarray}
where $k_b,k_b'$  denote boson momenta,
and $k_f$ denotes a fermion momentum.
These terms simply multiply the wavefunctions in the boundstate integral
equations.

\medskip

If we now include the contributions $F_h^{\dagger}
 \frac{1}{{\rm i}\partial_-} F_h$ in the light-cone Hamiltonian
(\ref{hamiltonian}), then we will 
encounter another type of singularity for vanishing longitudinal
momenta $k^+=0$. This singular behavior can be shown
to cancel a (divergent) momentum-dependent mass term, which is obtained
after normal ordering the  $F_h^{\dagger}\frac{1}{{\rm i}\partial_-} F_h$
interactions and performing an appropriate (infinite) 
renormalisation of the bare coupling $m_b$. This
 momentum-dependent mass term  
has the explicit form
\begin{eqnarray}
\lefteqn{ \frac{s^2 g^2 N}{2 \pi} \int_0^{\infty}
dk_1 dk_2 \left\{ \frac{}{} 
 \left( \frac{1}{k_2(k_1 - k_2)} + \frac{1}{k_2(k_1+k_2)} \right) 
\sum_h a_{h}^{\dagger}(k_1)a_{h}(k_1) \right.}  & & 
 \nonumber \\
& + & \frac{1}{k_2(k_1-k_2)} \sum_h 
b_{h}^{\dagger}(k_1)b_{h}(k_1)  
+  \frac{1}{k_2(k_1+k_2)} 
\sum_h 
d_{h}^{\dagger}(k_1)d_{h}(k_1)  \label{not2}
 \left.\frac{}{} \right\} 
\end{eqnarray}
The mechanism for cancellation here is different from the Coulombic case, since
we will require specific endpoint relations relating 
different wavefunctions.
Before outlining the general prescription for implementing
this cancellation, we consider 
a simple rendering of the boundstate integral equations involving
the $F_h^{\dagger}\frac{1}{{\rm i}\partial_-} F_h$ interactions.
In particular, let us consider the helicity zero 
sector with ${\cal N} =0$, and allow at most three partons.
Then the boundstate integral equation governing the behavior
of the wavefunction $f_{a_+ a_-}(k_1,k_2)$ for the two-boson state
$\frac{1}{N}\mbox{Tr}[a_+^{\dagger}(k_1)a_-^{\dagger}(k_2)]|0\rangle$ 
takes the form 
\begin{eqnarray}
M^2 f_{a_+ a_-}(x_1,x_2) & = & \frac{g^2 N}{\pi} \cdot 
        \frac{\pi}{4 \sqrt{x_1 x_2}}f_{a_+ a_-}(x_1,x_2)  
\nonumber \\
& + & \frac{s^2 g^2 N}{\pi} \sum_{i=1,2}\int_0^{\infty}
 dy \left( \frac{1}{y(x_i - y)} + \frac{1}{y(x_i+y)} \right) 
f_{a_+ a_-}(x_1,x_2) \label{nom} \\
& - & msg \sqrt{\frac{N}{2\pi}} \int_0^{\infty} d\alpha d\beta \hspace{1mm}
 \delta (\alpha + \beta - x_1) \times \nonumber \\
& & \frac{1}{\sqrt{x_1}} \left( \frac{1}{\alpha} + 
 \frac{1}{\beta} \right)\left[ f_{b_+ d_+ a_-}(\alpha,\beta,x_2) +
          f_{d_+ b_+ a_-}(\alpha,\beta,x_2) \right] + \dots\label{int1} 
\end{eqnarray}
where $M^2 = 2P^+P^-$, and $x_i = k_i/P^+$ are  (boost invariant)
longitudinal momentum fractions.
Evidently, the integral (\ref{int1})
arising from $1 \rightarrow 2$ parton interactions
behaves singularly for vanishing longitudinal momentum
fraction $\alpha \rightarrow 0$,
or $\beta \rightarrow 0$. However, these divergences
are  precisely canceled by the momentum-dependent mass terms
(\ref{nom}), which represent the contribution (\ref{not2}).

To see this, we may consider the integral equation governing
the wavefunction $f_{b_+ d_+ a_-}(k_1,k_2,k_3)$ for the 
three-parton state $\frac{1}{N^{3/2}} \mbox{Tr}[b_{+}^{\dagger}(k_1)
d_{+}^{\dagger}(k_2)a_-^{\dagger}(k_3)]|0\rangle$ :
\begin{eqnarray}
M^2 f_{b_+ d_+ a_-}(x_1,x_2,x_3) & = & m^2 \left( \frac{1}{x_1}
          + \frac{1}{x_2} \right)f_{b_+ d_+ a_-}(x_1,x_2,x_3) \nonumber \\
& + & \frac{g^2 N}{\pi} \sum_{i=1,2}
 \left[ \frac{1}{x_i} \left( \sqrt{1 + \frac{x_i}{x_3}} - 1 \right) \right]
 f_{b_+ d_+ a_-}(x_1,x_2,x_3) \nonumber \\
& - & msg \sqrt{\frac{N}{2\pi}}  \frac{1}{\sqrt{x_1+x_2}}
 \left( \frac{1}{x_1}
          + \frac{1}{x_2} \right) f_{a_+ a_-}(x_1+x_2,x_3) + \dots
\end{eqnarray}
If we now multiply both sides of the above equation by $x_i$,
and then let $x_i \rightarrow 0$ for $i=1,2$, we deduce the relations
\begin{equation}
f_{b_+ d_+ a_-}(0,x_2,x_3) = \frac{sg}{m}\sqrt{\frac{N}{2\pi}}
\frac{f_{a_+ a_-}(x_2, x_3)}{\sqrt{x_2}} 
\end{equation}
\begin{equation}
f_{b_+ d_+ a_-}(x_1,0,x_3) = \frac{sg}{m}\sqrt{\frac{N}{2\pi}}
\frac{f_{a_+ a_-}(x_1,x_3)}{\sqrt{x_1}} 
\end{equation}
It is now straightforward to show that the singular behavior
of the integral (\ref{int1}) involving the wavefunction
$f_{b_+ d_+ a_-}$ may be written in terms of a 
momentum-dependent
mass term involving the wavefunction $f_{a_+ a_-}$.
Similar divergent contributions are obtained from the  
the wavefunctions $f_{d_+ b_+ a_-}$, $f_{a_+ b_- d_- }$ and
$f_{a_+ d_- b_- }$, all of which may be re-expressed in terms of
the wavefunction $f_{a_+ a_-}$ by virtue of corresponding
`ladder relations'.
The sum of these divergent contributions exactly 
cancels the self-energy contribution (\ref{nom}).
An entirely analogous set of ladder relations
 were found for the case of fermions in the fundamental
representation of $SU(N)$ \cite{and96b}. 
 
For the general case where states are permitted to have more than
three partons, the correct ladder relations are not immediately
obvious from an analysis of the integral equations alone. 
Nevertheless, they
may be readily obtained from the constraint equation governing
the left-moving fermion field $\Psi_L$. In particular, we
have ${\rm i}\partial_- v_{\mp} = F_{\pm}$, and so 
vanishing fields at spatial infinity
would imply
\begin{equation}
          \int_{-\infty}^{\infty} dx^- F_{\pm}|\Psi \rangle = 0
\end{equation}
for color singlet states $|\Psi \rangle$. The analysis
of this condition in momentum space is quite
delicate, since it involves integrals of singular
wavefunctions over spaces of measure zero \cite{abd97}.
Viewed in this way we see that
the ladder relations are the continuum equivalent of zero mode constraint
equations that
have shown to lead to spontaneous symmetry breaking in discrete light-cone
quantization \cite{bep93}.

\section{exact solutions}
For the special case\footnote{ It should be remarked that
for 
any choice of Yukawa coupling $s$, {\em except} $s=0$,
the ladder relations 
are singular at $m=0$.
Therefore, in the context of the present 
analysis, we may only omit the fermion mass term 
if the Yukawa interactions are also omitted. Of course,
one may consider the limiting case $m \rightarrow 0$ 
for $s \neq 0$, but this is at present a numerically intractable
problem, and is perhaps better tackled using analytical considerations. 
}
 $s=t=m={\tilde m}_b = 0$, the only
surviving terms in the Hamiltonian 
(\ref{hamiltonian}) are the current-current interactions 
$J^+ \frac{1}{\partial_-^2} J^+$.
This
theory  has infinitely many massless boundstates, and  
the partons in these states are either
fermions or antifermions. 
States with bosonic $a_{\pm}$ quanta are always
massive. One also finds that the massless states are pure,
in the sense that the number of partons is a fixed integer,
and there is no mixing between sectors of different parton number.
In particular, for each integer $n \geq 2$, one can always find a massless
boundstate consisting of a superposition
of only $n$-parton states.  A striking feature
is that the wavefunctions of these states
are {\em constant}, and so these states are natural
generalizations of the constant wavefunction solution appearing in
t'Hooft's model \cite{tho74}. 

We present an explicit example below of such a constant
wavefunction solution involving a three fermion state with
total helicity $+\frac{3}{2}$, which is perhaps the simplest
case to study.
Massless states with five or more partons 
appear to have more than one wavefunction which are non-zero
and constant, and in general the wavefunctions are unequal.
It would be interesting to classify all states systematically, and
we leave this to future work. One can, however, easily
count the number of massless states. In particular,
 for ${\cal N }=3$, $h=+\frac{3}{2}$ states, there is
one three-parton state, $2$ five-parton
states, $14$ seven-parton
states and $106$ nine-parton states that yield massless solutions.

Let us now consider the action of the light-cone
Hamiltonian $P^-$ on the three-parton state 
\begin{equation}
|b_+ b_+ b_+ \rangle =
\int_0^{\infty} dk_1 dk_2 dk_3
\hspace{1mm} \delta (\sum_{i=1}^3 k_i - P^+) 
 f_{b_+ b_+ b_+}(k_1,k_2,k_3) 
\frac{1}{N^{3/2}} \mbox{Tr}[b_+^{\dagger}(k_1)
 b_+^{\dagger}(k_2)b_+^{\dagger}(k_3)]|0\rangle 
\end{equation}
The quantum number ${\cal N}$ is 3 in this case, and ensures
that the state $P^-|b_+ b_+ b_+ \rangle$ must have at least three
partons. In fact, one can deduce the following:
\begin{eqnarray}
\lefteqn{ P^- \mid b_+ b_+ b_+\rangle = 
\int_0^{\infty} dk_1 dk_2 dk_3
\hspace{1mm} \delta (\sum_{i=1}^3 k_i - P^+) \times} & & \nonumber \\
& & \left\{ -{g^2 N \over 2\pi} \right. \int^\infty_0 d\alpha d\beta
\frac{\delta (\alpha + \beta - k_1 -k_2)}
{(\alpha - k_1)^2} \left[ f_{b_+ b_+ b_+ }(\alpha,\beta,k_3)
- f_{b_+ b_+ b_+}(k_1,k_2,k_3) \right] \times \nonumber \\
& & \frac{1}{N^{3/2}}\mbox{Tr}
\left[b_+^\dagger(\alpha)b_+^\dagger(\beta)b_+^\dagger(k_3)\right]
\mid 0 \rangle  \nonumber \\
& &+{{g^2 N}\over 2\pi}\int^\infty_{0} 
d\alpha d\beta d\gamma \sum_h
{\delta(\alpha + \beta + \gamma - k_1) 
\over (\alpha+\beta)^2}
f_{b_+ b_+ b_+}(\alpha + \beta + \gamma ,k_2,k_3) \times \nonumber \\
& &\frac{1}{N^{5/2}}\mbox{Tr}\left[
\{b_h^\dagger(\alpha),d_{-h}^\dagger(\beta)\}b_+^\dagger(\gamma)
b_+^\dagger(k_2)b_+^\dagger(k_3)-
 \{b_h^\dagger(\alpha),d_{-h}^\dagger(\beta)\}
b_+^\dagger(k_2)b_+^\dagger(k_3)b_+^\dagger(\gamma)
\right] \mid 0 \rangle \nonumber \\
& &\nonumber \\
& &+{{g^2 N} \over 4\pi}\int^\infty_{0} d\alpha d\beta d\gamma
  \sum_h
{\delta(\alpha + \beta + \gamma - k_1 ) 
\over\sqrt{\alpha \beta}(\alpha+\beta)^2 }f_{b_+ b_+ b_+}
(\alpha + \beta + \gamma, k_2,k_3)
\nonumber \\
& & \frac{1}{N^{5/2}}\mbox{Tr}\left[
[a_h^\dagger(\alpha),a_{-h}^\dagger(\beta)]b_+^\dagger(\gamma)
b_+^\dagger(k_2)b_+^\dagger(k_3)-
 [a_h^\dagger(\alpha),a_{-h}^\dagger(\beta)] 
b_+^\dagger(k_2)b_+^\dagger(k_3)b_+^\dagger(\gamma)
\right] \mid 0 \rangle \nonumber \\
& + & \mbox{ cyclic permutations} \left. \frac{}{} \right\} 
\label{exacts}
\end{eqnarray}
The five-parton states  above
correspond to virtual fermion-antifermion and boson-boson
pair creation.
The expression (\ref{exacts}) vanishes if the wavefunction
$f_{b_+ b_+ b_+}$ is constant.

\section{Numerical Results}
One may perform a numerical analysis of the boundstate integral equations 
by discretising 
the longitudinal momentum fractions $x_i$ so
that $x_i =1/K,2/K,3/K,\dots$,
where $K$ is some (ideally) large positive integer.
The problem is then reduced to finite matrix diagonalisation \cite{pab85}.

In this initial investigation, we are primarily interested in the
qualitative behavior of the theory, and so we make
a choice of couplings which we believe summarizes
the general properties of the theory. 
A more detailed study of the coupling-constant
space of the theory will be presented elsewhere
\cite{anp97}.
In Fig. 1(a) we have plotted the mass spectrum for the baryon-like 
sector ${\cal N}=3$
with the special choice of couplings $t=s=m={\tilde m}_b=0$.
The horizontal axes $<n>$ represents the average number of partons 
in a boundstate. Pure massless states of length $3$, $5$, $7$,$\dots$
are found, and there is in general a high level of degeneracy.
Massless states of length $2$, $4$, $6$,$\dots$ appear in the
meson-like sector ${\cal N}=2$, and the associated spectrum
is qualitatively the same. In particular, one finds a concentration
of states with average length $<n>$ nearly an integer.
 
In Fig. 1(b) we have allowed the coupling constants $s,t$ and $m$
to be non-zero. There is now a lower trajectory of
states concentrated near $<n>=3,4,5,6,\dots$. The presence of
Yukawa interactions means there are $1 \leftrightarrow 2$ parton
processes which flip the sign of the helicity for a fermion
or antifermion. This 
dramatically reduces the high degeneracy exhibited in Fig. 1(a). 

These `helicity-flip' terms are responsible for the
complicated dynamics governing the polarized distribution
of fermions and bosons in a given boundstate. 
Figs 1(c) and 1(d) represent plots
of the polarized and unpolarized fermion and boson structure functions
for the lightest baryon-like boundstate appearing in Fig. 1(b).
These curves are normalized such that the area under the 
polarized quark structure function {\em plus} the area under
the polarized gluon structure function is equal to the 
total (conserved) helicity, which is $+\frac{1}{2}$ in
this case. Evidently, a better understanding of the small-$x$ behaviour
of these structure functions can only be achieved if larger values
of $K$ are investigated, since one expects sea-quarks and small-$x$
gluons (`wee partons') to contribute significantly to the overall
polarization of a boundstate.

Finally, in Figs 1(e) and 1(f) we provide evidence for 
a phase transition of the field theory if the quartic coupling
$t$ is sufficiently large and negative. In particular, we observe a 
regime where the theory sharply becomes tachyonic, and the 
groundstate has diverging average length $<n>$. This behavior
is entirely consistent with earlier work on the 
phase transitions of large-$N$ matrix models
with cubic and quartic interactions \cite{and95}.

\section{ Conclusions }

We have presented a non-perturbative Hamiltonian formulation of a class 
of $1+1$ dimensional matrix field theories, which may be derived 
from a classical dimensional reduction of 
${\mbox{QCD}}_{3+1}$
coupled to Dirac adjoint fermions. We choose
to adopt the light-cone gauge $A_- = 0$, and are able
to solve numerically the boundstate
integral equations in the large-$N$ limit.
Different states may be classified according to total
helicity $h$, and the quantum number ${\cal N}$, which
defines the number of fermions minus the number of antifermions
in a state. 

For a special
choice of couplings that eliminates all interactions
except those involving the longitudinal current $J^+$, 
we find an 
infinite number of pure massless states of arbitrary length.
The wavefunctions of these states are 
always constant, and may be solved for
exactly. An example was presented in Section IV.
In general, a massless solution involves several
(possibly different) constant wavefunctions. 
We anticipate that the massless solutions observed in
studies of $1+1$ dimensional
supersymmetric field theories \cite{maa95}  are analogous to the 
constant wavefunction solutions found here.

When one includes the Yukawa interactions,
singularities at vanishing longitudinal momenta  
arise, 
and we show in a simple case how these are canceled 
by the boson and fermion self-energies.
This cancellation relies on the derivation of certain
`ladder relations', which relate different 
wavefunctions at vanishing longitudinal momenta.
These relations become singular for vanishing fermion mass
$m$, and so in the context of the numerical techniques 
employed here, one is prevented from studying the limit
$m \rightarrow 0$. Analytical techniques which are 
currently under investigation are expected to be relevant
in this limiting case \cite{abd97}.

In the Figures 1(a)-(f), we have summarized some of
the generic features exhibited by these matrix models.
The plot of spectra suggests that the density of states grows 
with average length. However, larger values of the cutoff $K$
will be necessary before we can draw any firm conclusions.

A particularly important property of these models is that 
virtual pair creation and annihilation of bosons and fermions
is not suppressed in the large-$N$ limit, and so our
results go beyond the valence quark (or quenched) approximation.
This provides the scope for strictly field-theoretic 
investigations of the internal structure of boundstates where
 `sea-quarks' and small-$x$ gluons are expected to contribute
significantly to the overall polarization of a boundstate.  
In this work, we were able to calculate 
structure functions for the baryon-like and meson-like
states, but only for relatively low values of $K$. The 
baryon-like state with total helicity $h=+\frac{1}{2}$ perhaps
best resembles a nucleon, and so we provide plots of spectra
and structure functions in Fig.1(a)-(d). The most important
aspect of these structure functions is perhaps the behavior at 
small-$x$, but much larger values of $K$ will be necessary
in order to probe this region numerically.

The techniques employed here are not specific to the 
choice of field theory, and are expected to have a
wide range of applicability, particularly in the 
light-cone Hamiltonian formulation of supersymmetric
field theories.

\acknowledgments
\noindent
The work was supported in part by a grant from
the US Department of Energy. Travel support was provided in part by a NATO
collaborative grant.

\newpage
\begin{center}
FIGURE CAPTIONS
\end{center}

\noindent
Fig.1(a) Mass spectrum (in units $g^2N/\pi$) for the 
baryon-like sector;
 ${\cal N} = 3$, $h=+\frac{1}{2}$, ${\tilde m}_b = m = t = s = 0$,
 $K=7$ (465 states). 

\medskip

\noindent
Fig.1(b) Mass spectrum (in units $g^2N/\pi$)
 for the baryon-like sector with general
 couplings;  ${\cal N} = 3$, $h=+\frac{1}{2}$, ${\tilde m}_b = 0$,
$m^2/(g^2 N/\pi)=1$, $t=-0.1$, $s = 0.5$,
 $K=6$ (285 states).

\medskip

\noindent
Fig.1(c) Polarized and unpolarized (upper curve) fermion 
structure function for the lightest baryon-like state in Fig. 1(b).
The area under the lower curve gives the contribution of the
total helicity coming from the fermions and antifermions.

\medskip

\noindent
Fig.1(d) Polarized and unpolarized (upper curve) boson 
structure function for the lightest baryon-like state in Fig. 1(b).
The area under the lower curve gives the contribution of the
total helicity coming from the bosons.

\medskip

\noindent
Fig.1(e) Mass (in units $g^2N/\pi$) of the groundstate
 in the sector ${\cal N} = 0$, $h=+2$, for variable quartic coupling
 $t$ (${\tilde m}_b = m = s = 0$, $K=5$).

\medskip

\noindent
Fig.1(f) Average number of partons $<n>$ in the groundstate
(of Fig. 1(e))
 for variable quartic coupling $t$.   

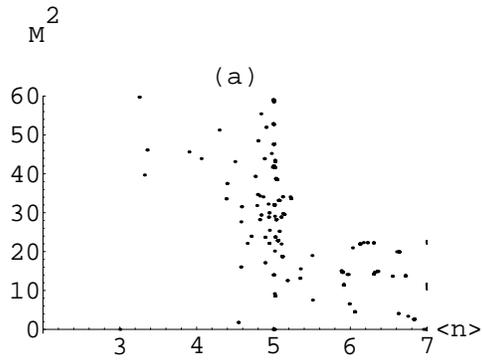
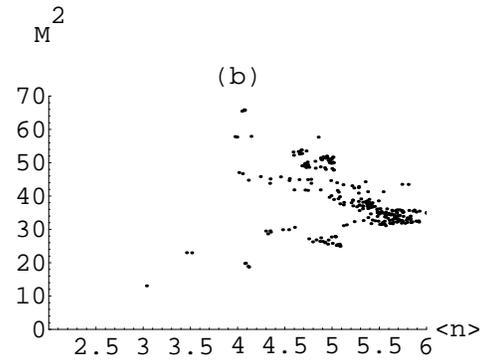
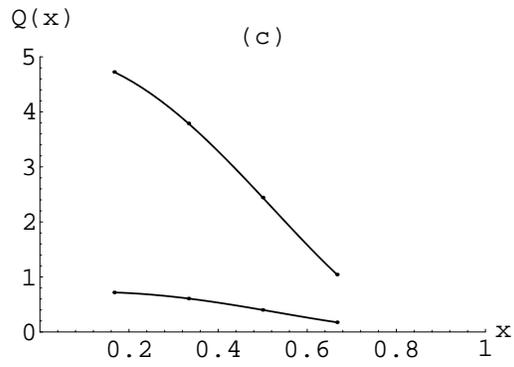
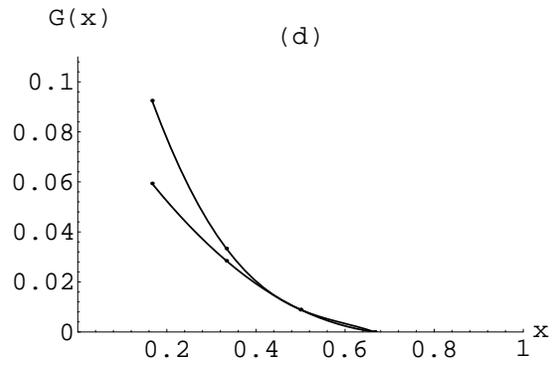
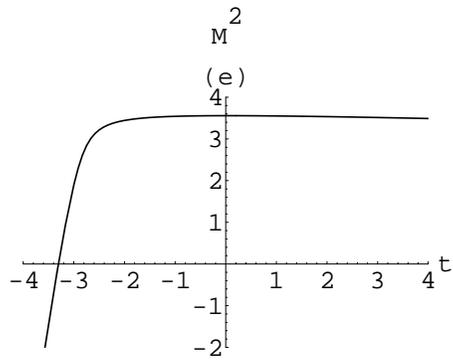
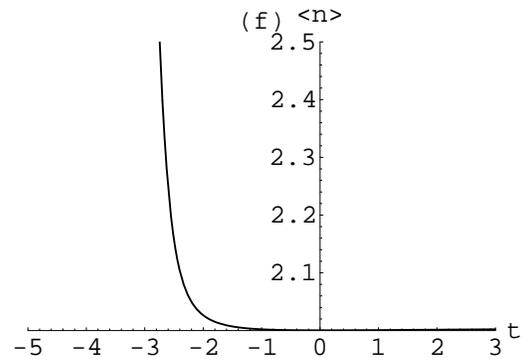
\end{document}